# Phase memory for optical vortex beams

**MAHDI ESHAGHI, CRISTIAN HERNANDO ACEVEDO, MAHED BATARSEH, ARISTIDE DOGARIU**

*CREOL, The College of Optics and Photonics at the University of Central Florida, 4304 Scorpius St, Orlando, Orlando, Florida 32816, USA.*

**Optical vortex beams have been under considerable attention recently due to their demonstrated potential for applications ranging from optical communication to particle trapping. Practical problems related to the dependence between their phase structure and the physical size have been addressed by introducing the concept of perfect optical vortex beams. Propagation of these structured beams through different levels of disturbances is critical for their uses. For the first time, we examine quantitatively the degradation of perfect optical vortex beams after their interaction with localized random media. We developed an analytical model that describes how the spatial correlation length and the phase variance of the disturbance affect the phase distribution across the vortex beams. This allows to ascertain the regimes of randomness where the beams maintain the memory of their initial vorticity. Systematic numerical simulations and controlled experiments demonstrate the extent of this memory effect for beams with different vorticity indices.**

## Introduction

It has been shown before that, in free space propagation, the electromagnetic field could contain robust phase singularities. In these points the amplitude vanishes and the phase cannot be determined: a field property which was termed "screw dislocations" due to the similarity to crystal lattice defects [1, 2]. When the wave propagates, the lines of constant phases around a singularity trace out a spiral that is mathematically similar to superfluid vortices, which inspired the term "optical vortex" [3]. To some degree, the vortices embedded in a light beam, termed "optical vortex" beams (OV), act as charged particles so that they may rotate around the beam axis, repel and attract each other, and created or annihilated in dipole pairs [4, 5].

Since the beginning, the properties of optical vortices attracted a significant attention, especially in the case of random fields [6, 7, 8], where it has been shown that the number density of vortices can be pretty high. In fact, in the case of so-called fully developed speckle patterns, one optical vortex (zero amplitude) accompanies one speckle spot (maximum amplitude) [6, 9, 10, 11]. It is worth mentioning that vortices are characterized by their "singularity strength" or "topological charge" (TC), which is the signed sum of local vortices inside a determined closed contour [12].

Interferometric techniques are generally used to detect the phase singularities where "spiral" and "fork" shaped fringes are observed in the cases of collinear and tilted interference, respectively [7, 13]. Non-interferometric techniques are also used by taking advantage of far-field diffraction of OVs by different elements [14, 15]. Note that all these methods determine the TC of a singularity without any additional information about the spatial distribution of different vortices. This problem can be addressed by simultaneous measurement of the beam amplitude and phase [16, 17].

The light field of an OV carries angular momentum [18, 19, 20], including orbital (OAM) and spin (SAM) parts, which arises from the light surrounding the singularity because the line of singularity does not carry energy and it has no momentum [21]. In the case of OVs embedded into azimuthally symmetric beams, such as Bessel [22, 23], Laguerre-Gauss [24], and Bessel-Gauss [25], the TC and OAM share the same value. However, there are many other cases where the values of these two parameters differ [12, 16, 26]. Therefore, one can conclude that there is no general relationship between these two parameters.

Several reports conjectured that possessing OAM may help OVs propagate through optical turbulence with less distortion than conventional Gaussian beams [27, 28, 29]. Other experimental studies have shown that in random aerosol media the OVs had mostly lower stability at small propagation distances than a Gaussian beam. In contrast, at considerable distances, OVs have been found to be more resilient [30]. It is worth noting that, indeed, there are different criteria for beam stability. Sometimes maintaining TC is a crucial characteristic while in other situations the so-called scintillation index can be an appropriate parameter. It has been shown that the TC is a robust quantity since it can be transmitted over significant distance in the presence of atmospheric turbulence without any loss, which means it can be utilized to encode information in free-space optical communications [31]. On the other hand, the OAM modes span an infinite-dimensional basis, which can be used to send photon-level information [32]. Nevertheless, after passing through different kinds of turbulent atmospheres, the core of a vortex beam wanders away from its original, which seriously hinders optical communications [31, 33]. Describing the statistical properties of the phase dislocations after the beam has been disturbed is critical and, therefore, the topic received considerable attention in recent years [27, 33, 34]. A few works addressed the possibility to recover the vorticity of OV beams after scattering by random phase screens [35, 36, 37]. However, most of these reports are rather qualitative and without a clear description of what is meant by "diffuser" or "ground glass."

Here we provide a quantitative description of the OV degradation due to interaction with inhomogeneous media, which is statistically characterized by the "spatial correlation length" and the "variance" of phase randomness". In particular, we address the case of a "perfect optical vortex beam" (POV) for which the size of the beam is independent of its phase structure [38, 39]. Such beams are Fourier transforms of the Bessel beams [40] and can be created by axicon [41], width-pulse approximation of the Bessel function [39], liquid crystal spatial light modulators (SLMs) [42], and digital micro-mirror devices (DMDs) [43]. Although the information capacity is inherently reduced, their use in free space and underwater communications has been recently demonstrated [44, 45, 46, 47]. In the following, we will develop the exact statistical relation between the TC and the OAM modes describing the perturbed field. On this basis, we will then quantify the extent of vortex memory, i.e., the range of randomness for which the initial vorticity can still be recovered. This memory range will be first established theoretically and then demonstrated experimentally.

## Theoretical model

Let us start from the description of a Bessel-Gauss beam [25]
$$E_1(\rho, \phi) = J_m(k_r\rho) \, e^{im\phi} \, e^{-\rho^2/\omega_g^2}, \quad (1)$$
where $m$ is the TC, $J_m$ is the Bessel function of the first kind of order $m$, and $\omega_g$ is beam waist of the Gaussian beam used to generate the Bessel beam (Appendix 1). The Fourier transformation of $E_1(\rho, \phi)$ leads to the far-field distribution
$$E_2(r, \theta) = i^{m-1}(\omega_g/\omega_0)e^{im\theta}e^{-(r^2+r_0^2)/\omega_0^2}I_m(2r_0r/\omega_0^2) \quad (2)$$
where $I_m$ is the modified Bessel function of the first kind of order $m$, $2\omega_0$ and $r_0$ are the ring width and radius, respectively. When the argument of $I_m$ is sufficiently large, this function behaves as an exponential, and Eq. 2 describes a so-called POV beam [40]
$$E_2(r, \theta) \cong E_0 e^{im\theta}\delta(r - r_0) \quad (3)$$
where $E_0 = i^{m-1}(\omega_g/\omega_0)$ is a constant complex coefficient. This beam is incident on a random phase screen $T(r, \theta) = exp(i\psi(r, \theta))$ where $\psi(r, \theta) = \psi_0 + \psi'(r, \theta)$ is a random function of mean $\psi_0$. The probability distribution will be described later. At a distance $z$ in the far-field, the distribution of the perturbed field can be written in polar coordinates as
$$E_3(\rho, \phi) = (E_0/N) \sum_{n=1}^{N} e^{i\psi'_{(n;r=r_0)}} \, e^{iR_n} \quad (4)$$
where $R_n = \psi_0 + m\theta_n - \mu r_0\rho cos(\phi - \theta_n)$ is a deterministic function of $n$ and $\mu = \pi/\lambda z$. Along the POV circumference, $N$ represents the number of independent random phase elements such that $\theta_n = n(2\pi/N)$ for $n \in [1, N]$. For the sake of notation brevity, we will replace $\psi'_{(n;r=r_0)}$ by $\psi_n$ and $E_3(\rho, \phi)$ by $E(\rho, \phi)$. In describing the random phase screen, we will define a "spatial correlation length" that encompasses $d$ elements with the same phase. When $N \gg d$, this correlation length is spatially invariant along the circumference of the ring. We will also consider a Gaussian distribution for the probability density function $p(\psi')$ of random phases over an adjustable range $[-\alpha\pi, \alpha\pi]$. Thus, the normalized probability density function is $p(\psi') = \left(\sqrt{2\pi s^2} \, erf(\alpha\pi/\sqrt{2s^2})\right)^{-1} exp(-\psi'^2/2s^2)$ where $s$ is the standard deviation, and $erf$ denotes the error function. Choosing this distribution let one probe the whole range between *Dirac delta function* ($s \to 0$) toward *uniform* ($s \to \infty$) distributions with different possible ranges of phases which is determined by parameter $\alpha$. Throughout this manuscript, we refer to $\alpha$ as "phase variance."

As detailed in Appendix 2, the field $E(\rho, \phi)$ can be expanded on a vorticity basis like
$$E(\rho, \phi) = \sum_{k=-\infty}^{\infty} H(k; \rho) \, e^{ik\phi} \quad (5)$$
where each complex coefficient $H(k; \rho)$ is defined as
$$H(k; \rho) = \Phi. \, A \sum_{n=1}^{N/d} e^{i\psi'_n} e^{i[m-k]\theta_{(n-1)d}} \quad (6)$$
with $\Phi = (-i)^{|k|}e^{i[m-k]\frac{d+1}{N}\pi + i\psi_0}$ and $A = J_{|k|}(\mu r_0\rho)sinc([m-k](d/N)\pi)$ are phase and amplitude terms that factorize out of the summation, including the random phase terms. The case of no randomness means either $N = d$ or $\psi_n = 0$ for $n \in [1, N/d]$. The first one is a trivial case of uniform phase, while in the second case, one can easily show that Eq. (5) becomes $E(\rho, \phi) = E_0(-i)^{|m|}J_{|m|}(\mu r_0\rho)e^{im\phi}$ where it is evident that initial vorticity is preserved. Based on the distribution of $p(\psi')$, one can find the statistical properties of each complex coefficient $H(k; \rho)$, including the average and all other higher-order moments. In the following, we will study in detail the behavior of the effective TC and the OAM modal decomposition of $E(\rho, \phi)$ in response to changes in the randomness defined by its correlation length and phase variance. To this end, we will evaluate the effective TC on a circular contour of arbitrary radius $\rho_0$ which is centered on the optical axis. Inside this contour, we can calculate the weights $V^k = \langle|H(k; \rho)|^2\rangle|_{\rho=\rho_0}$ for vorticity of order $k$ [48]. The corresponding OAM mode can then be gauged as $L^k = \int_0^{\rho_0}\langle|H(k; \rho)|^2\rangle\rho \, d\rho$ [49]. The most probable values for the vorticity and OAM modes are further determined by evaluating the statistical average $\langle x \rangle = \sum_i p_i x_i / \sum_i p_i$ where $x$ can be either $V$ or $L$. A step-by-step derivation of these parameters is included in Appendix 3 together with the closed analytical form for the average of each TC and OAM modes inside the circular contour.

The extent of "vortex memory" or "phase memory" is determined by the rate at which the initial vorticity $m$ vanished when changing the randomness parameters. A quantitative criterion can be established by defining, for instance, the regime of randomness for which the average vorticity remains larger than $m - 1$. Up to this point, the field has maintained a certain memory of its initial conditions such that a "global phase" can be used to recover the initial vorticity. With increasing disturbance, the mean vorticity continues to decay from $m - 1$ to 0 because the perturbed field evolves towards a stochastically homogenous and isotropic field (speckle pattern) in which case the only meaningful property will be the "local phase" distribution as shown in Appendix 4.

## Results and Discussion

The overall goal is to understand how the number $|m|$ local vortices evolve after a beam with given initial vorticity is affected by the interaction with the random phase screen [5]. For this, one needs to define a suitable observation scale such that vortices close to the optical axis can be tracked appropriately. It is known that the mean number density of dislocations inside a fully developed speckle pattern equals $1/2a_{coh}$ where $a_{coh} = \lambda^2 z^2/\pi r_0^2$ is the average coherence area [10]. This means that, within a circle of radius $\rho_0 = \sqrt{a_{coh}/\pi}$ such that $\mu r_0\rho = 1$, there are, on average, two vortices of opposite handedness. In the following, we will use $\rho_0$ as to limit the spatial extent of our analysis. This choice is also convenient practically because a topological charge larger than unity within a circle of radius $\rho_0$ indicates locally high-order vorticity or a "global phase" that can be easily determined without measuring the spatial distribution of the complex field.

We conducted a detailed numerical simulation of different interaction regimes. After generating the stochastic field based on the initial beam structure and randomness properties, we implemented an algorithm to find the location of vortices, as well as their handedness (see Appendix 5). Then by changing the randomness parameters, we track the position of main vortices generated because of initial vorticity, and finally find the total charge inside the chosen closed contour. We have also conducted an experimental demonstration where a "tunable phase screen" was used to probe a range of randomness parameters. The phase measurement over the random optical field was then conducted in less than $1ms$ using appropriate retardance modulators. The vorticity was then assessed by summing over the phase gradient along the circumference of closed contours as detailed in the Appendix 6.

The quantitative comparison between the analytical, numerical, and experimental results is illustrated in Fig. 1. Two important conclusions emerge from these results. The first one relates to the influence of the spatial correlation length of randomness. Its effect on the effective topological charge is not linear, and randomness with higher spatial correlation distorts the phase structure more dramatically. In the limit $d \to 0$ one can see that the lifetime of the initial vorticity varies linearly with increasing of the correlation length. Second, under similar conditions of phase disturbance, fields with higher vorticity are decorrelated faster. This means that when the number of bunched local vortices with the same handedness increases, they tend to wander more rapidly. This behavior can be explained by a simple analogy where local vortices are seen as interacting particles with positive (or negative) charges according to their handedness. The same sign charges repel each other stronger when more of them are gathered within a finite space.

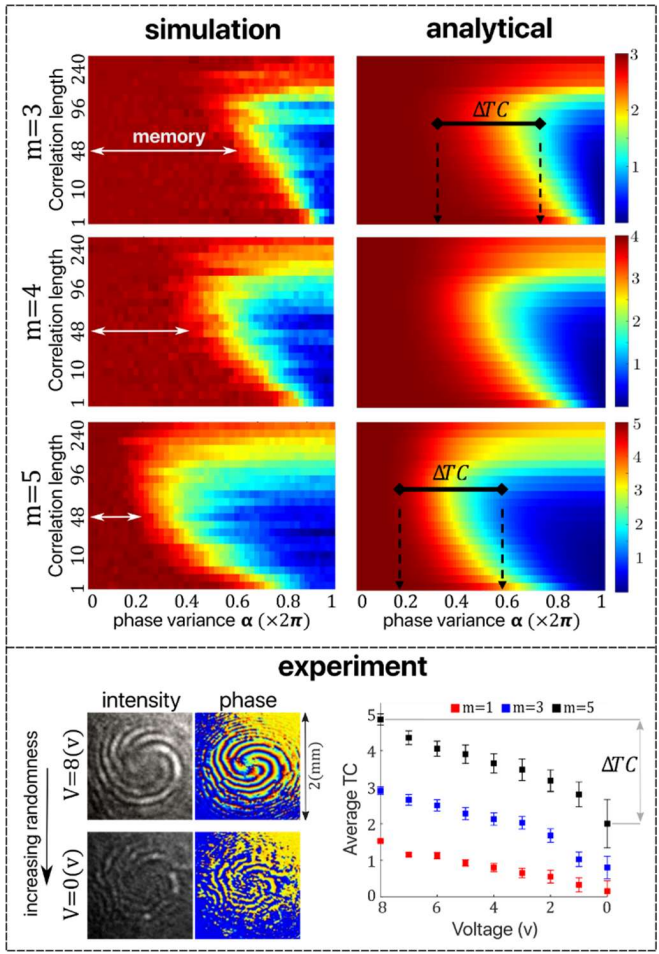

Fig. 1. Analytical, numerical, and experimental data regarding the phase memory in optical vortex beams. The white arrows denote the range of phase variance where the initial vorticity can be recovered, i.e. the extent of "vortex memory". The numerical simulation data reflects an average over 25 realizations of randomness. The $\Delta TC$ denotes the experimental range of variation for the randomness parameters in the case of $m = 3$ and $m = 5$. Details regarding the numerical simulations and the experiment are included in the supplementary materials.

A significant consequence of our analytical result is that it establishes a relationship between the average weight of each OAM mode and the vorticity characterizing the final random field. In general, there is no deterministic relation between these two properties of inhomogeneous fields but here we have derived a clear statistical connection between them. Thus, statistical characteristics of one property can be evaluated for measurements of the other, which could be useful, for instance, in optical communication where proper measurements of OAM modes are desirable.

Lastly, our results are directly relevant to the associated stochastic inverse problem. When the characteristics of random media can be appropriately modeled as space-variant phase-only perturbations, the statistical relations we established between the randomness parameters and the effective vorticity can enable simple sensing strategies where POVs with different vorticities are used to interrogate the randomness. Measurements of global phases in the corresponding inhomogeneous fields could then be used to determine the variance and the spatial correlation of the phase disturbance.